\documentclass{aa}
\usepackage{graphicx}
\usepackage{txfonts}

\begin{document}

   \title{Dynamical feedback of the curvature drift instability on its saturation process}

   \author{Z. Osmanov\inst{1}\thanks{E-mail:
z.osmanov@astro-ge.org (ZO); d.shapakidze@astro-ge.org (DS);
g.machabeli@astro-ge.org (GM) } D. Shapakidze$^{1}$ and G.
Machabeli$^{1,2}$
          }

   \offprints{Z. Osmanov}

   \institute{E. Kharadze Georgian National Astrophysical
   Observatory, Chavchavadze State University,
              Kazbegi str. $2^a$, Tbilisi, 0106, Georgia
         \and
   Javakhishvili State University, Chavchavadze str. $3$, Tbilisi,
   0179, Georgia
             }



  \abstract
  {}
   {We investigate the reconstruction of pulsar magnetospheres close
   to the light cylinder surface to study the curvature drift instability
(CDI) responsible for the twisting of magnetic field lines in the
mentioned zone. The influence of plasma dynamics on the saturation
process of the CDI is studied.}
   {On the basis of the Euler, continuity, and induction
equations, we derive the increment of the CDI and analyze
parametrically excited drift modes. The dynamics of the
reconstruction of the pulsar magnetosphere is studied analytically.}
   {We show that there is a possibility of a parametrically excited
rotational-energy pumping-process in the drift modes. It is
indicated by the generation of a toroidal component of the magnetic
field that transforms the field lines into such a configuration, in
which plasma particles do not experience any forces. At this stage,
the instability process saturates and the further amplification of
the toroidal component to the magnetic field lines is suspended.}
   {}


   \keywords{instabilities - plasmas - pulsars: general - acceleration of particles}

   \maketitle
%


\section{Introduction}

Exploring the magnetic field of the Crab nebula, \cite{pid} was the
first to discover the presence of a central object in the nebula,
with frozen magnetic field inside. It was assumed that the rotation
of a central object provokes the generation of the toroidal
component of magnetic field. Investigations have also shown that
this kind of magnetic field characterizes magnetized star outflows
(\cite{weber}). Pulsars are one of the most distinctive examples of
rotationally powered magnetized stars producing the relativistic
outflows commonly known as pulsar winds.

One of the major problems for pulsar winds concerns the transition
of the magnetized plasma flows by means of the so-called "light
cylinder" surface, a hypothetical zone, where the linear velocity of
rotation equals the speed of light. According to the model of pulsar
magnetosphere (\cite{gj}), the relativistic plasma flow, emanating
from the pulsar surface streams along very strong magnetic field
lines. The typical values of the magnetic field are of the order of
$10^{12}$ G (\cite{manch}). Such a strong magnetic field forces the
particles, streaming toward the light cylinder zone, to co-rotate
with a pulsar. On the other hand, once the relativistic effect of
the mass increment is taken into account, the radial acceleration of
the particles appears to be limited. This means that the particles
will never cross the light cylinder surface if the rigid rotation is
preserved (\cite{mr}). Therefore, the co-rotation of the magnetized
plasma cannot be maintained close to the mentioned zone and a
further exploration of this phenomenon is needed.

The simplest means of explaining the transition in the plasma
outflow at the light cylinder surface is to constrain the motion of
the flow to become asymptotically close to a regime, in which
particles do not experience any forces (see e.g.,
\cite{mmsh2000,r03}). For this purpose, if the magnetic field is
still strong, one needs to twist the field lines in an appropriate
way. In other words, it is necessary to generate the toroidal
component of the magnetic field. The pulsars maintain the dipolar
magnetic field with a certain curvature of the field lines. The
particles moving along the curved field lines drift perpendicularly
to the curvature plane. The existence of the drift motion creates
the necessary conditions for developing a CDI (see e.g.,
\cite{kmm89,kmm91a,kmm91b,shmmkh03}). The mechanism excites
transverse drift modes providing the magnetic field with the
toroidal component. The CDI becomes effective in the vicinity of the
light cylinder surface, where the pulsar's dipolar field is
comparable to the toroidal component of the excited mode. The
kinetic energy of the primary beam particles, of Goldreich-Julian
(GJ) density, can indeed supply the generation of the drift waves.
Mostly, it is assumed that the toroidal component of the magnetic
field is created by the GJ current. However, the kinetic energy
density, ${W}_b$, of the primary beam is insufficient to change the
configuration of the dipolar magnetic field, $B_0$, significantly,
since ${W}_b<<B_0^2/(4\pi)$. Therefore, the process of generating
the toroidal component, $B_r$ (of the order of $B_0$), requires an
additional energy supply, which can be extracted from the pulsar
rotational energy. We have already shown that the rotational energy
pumping into the drift modes can be implemented by the "parametric"
instability (\cite{mnras}). This mechanism is called parametric,
because the effect is caused by the relativistic centrifugal force,
which, as a parameter, changes in time and induces the instability.
The aim of the present paper is to study the saturation process of
this instability close to the light cylinder zone.

The problem analysis is completed well if one considers the
mechanical analog of the motion of plasma flow along the pulsar
magnetic field lines and studies the dynamics of a single particle
moving inside the rotating channels. In the present paper, we apply
the method developed in (\cite{r03}) and study the plasma process
that converts the configuration of the magnetic field lines to the
Archimedes' spiral: it appears that in order to suppress the
reaction force, the field lines, rather being straight, should
deviate back and lag behind the rotation; consequently, the
curvature will eventually increase, inevitably leading to the
decrement of the reaction force; this process will last until the
magnetic field lines form the shape of the Archimedes' spiral, which
removes the reaction force and saturates the instability.

The paper is arranged as follows. In section 2, we examine the
parametric curvature drift instability and derive an expression for
its corresponding growth rate. In section 3, we consider the
kinematics of a particle moving along the rotating magnetic field
line. In section 4, we apply the method for 1-second pulsars and the
Crab pulsar, and in section 5 we summarize our results.

\begin{figure}
 \par\noindent
 {\begin{minipage}{0.35\linewidth}
 \includegraphics[width=\textwidth] {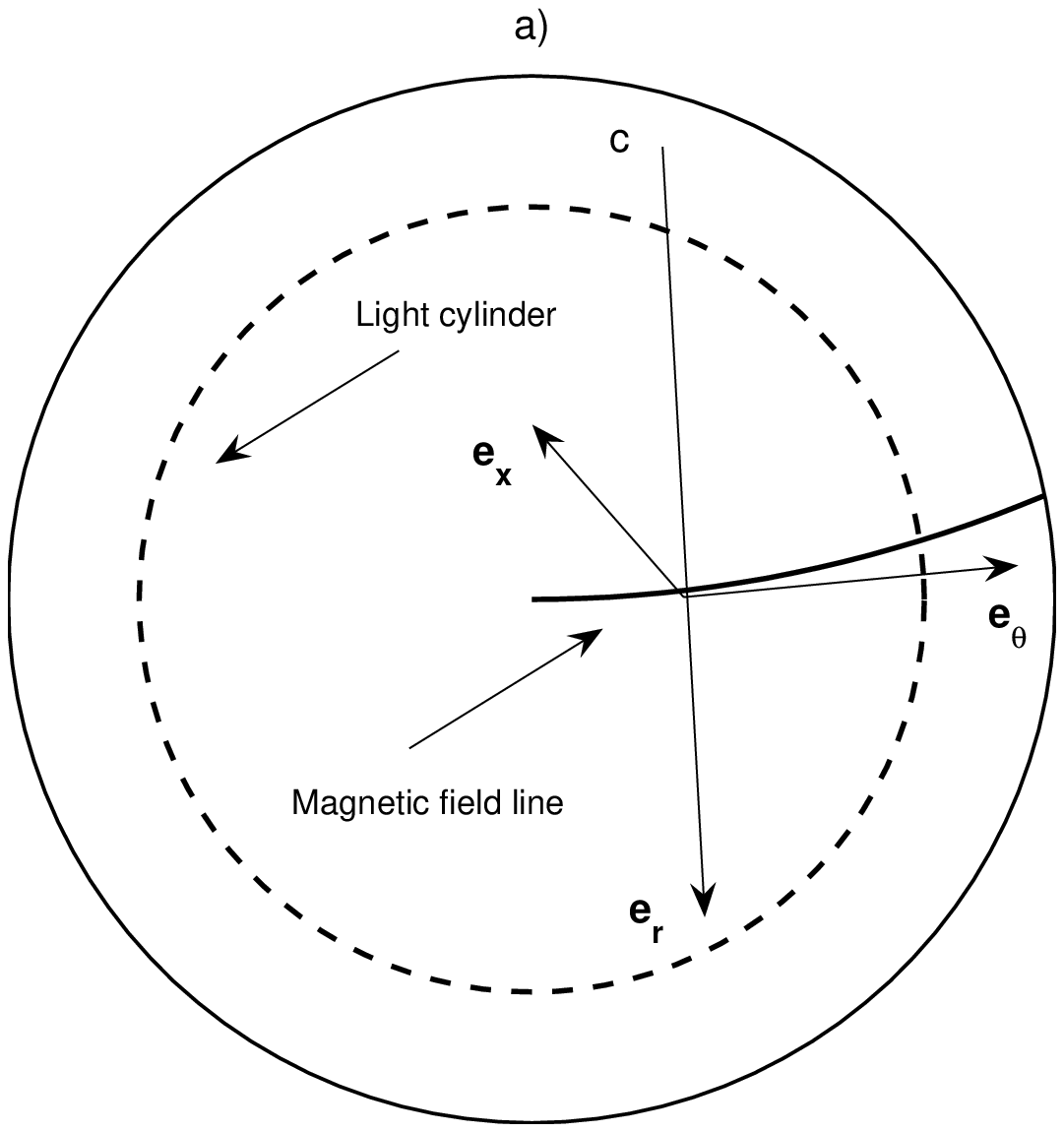}
 \end{minipage}
 }
 \hfill
 {\begin{minipage}{0.35\linewidth}
 \includegraphics[width=\textwidth] {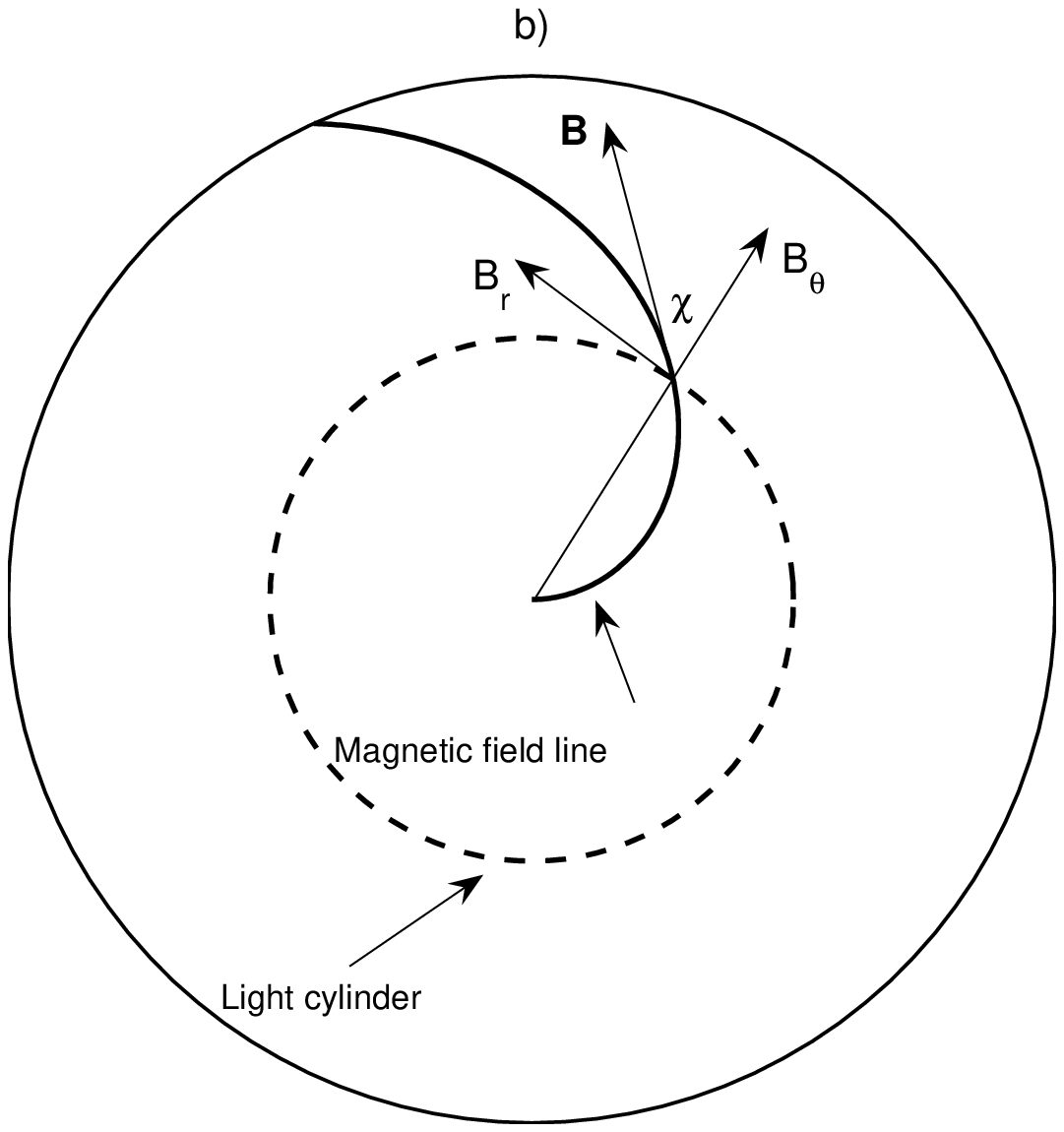}
 \end{minipage}
 }
 \caption[ ] {\footnotesize $a$) The geometry in
which the set of main Eqs. (\ref{eul}-\ref{ind}) is considered;
(${\bf e}_{\theta},{\bf e}_r,{\bf e}_x$) denotes orthonormal basis
of unit vectors; ${\bf e}_x$ is directed perpendicularly to the
plane of the figure; $C$ is the center of the curvature; and ${\bf
e}_{\theta}$ and ${\bf e}_r$ are the tangential and perpendicular
(with respect to the field line) unit vectors respectively. $b$) The
geometry for deriving Eq. (\ref{time}): the curved line denotes the
twisted magnetic field, ${\bf B}$, generated due to the raising of
magnetic perturbation, ${\bf B}_r$. Note, that ${\bf B}_r$ and ${\bf
B}_{\theta}$ are oriented with respect to the initial `quasi
straight' magnetic field line. }\label{fig3}
 \end{figure}

%
\section{Parametric curvature drift instability}
%

It is well known that the presence of an external varying parameter
usually generates the plasma instability. For example, the mechanism
of energy pumping process from the external alternating electric
field into the electron-ion plasma is quite well investigated (see
e.g., \cite{silin1,gal,max}). Although the physics of the parametric
instability in the electron-ion plasma differs from that of the
electron-positron ($e^{-}e^{+}$) plasma, the techniques of
calculation can be the same. The case of a $e^{-}e^{+}$ plasma,
where the external varying parameter is the altering centrifugal
acceleration, was considered by Machabeli et al.
(2005). 

We start our consideration by initially assuming that the magnetic
field lines are almost straight with very small non zero curvature
(see Fig. \ref{fig3}a). The plasma stream is supposed to move along
the co-rotating field lines of a very strong magnetic field ($\sim
10^{12}$ G for typical pulsars). We assume that the plasma flow
consists of two components: the plasma component composed of
electrons and positrons ($e^{\pm}$); and, the so-called, beam
component ($b$) composed of ultra relativistic electrons. The
dynamics of plasma particles moving along the straight rotating
magnetic field lines is described by the Euler equation
(\cite{incr1})

\begin{equation}
\label{eul} \frac{\partial{\bf p_{\alpha}}}{\partial t}+({\bf
v_{\alpha}\nabla)p_{\alpha}}=
-c^2\gamma_{\alpha}\xi{\bf\nabla}\xi+\frac{e_{\alpha}}{m}\left({\bf
E}+ \frac{1}{c}\bf v_{\alpha}\times\bf B\right),
\end{equation}
where $$ \label{xi} \xi\equiv \sqrt{1-\Omega^2R^2/c^2}\,, $$ $R$ is
the coordinate along the straight field lines,
$\alpha=\{e^{\pm},b\}$ denotes the sort of particles, ${\bf
p_{\alpha}}$, ${\bf v_{\alpha}}$, and  $e_{\alpha}$ are the momentum
(normalized to the particle's mass), the velocity, and the charge of
electrons/positrons, respectively; ${\bf E}$ is the electric field
and ${\bf B}$ is the magnetic field. On the right-hand side of Eq.
(\ref{eul}), there are two major terms, the first of which
represents the centrifugal force and the second, the Lorentz force.
The full set of equations for $n,{\bf v},{\bf E}$, and ${\bf B}$
should be completed by the continuity equation
\begin{equation}
\label{cont} \frac{\partial n_{\alpha}}{\partial t}+{\bf
\nabla}(n_{\alpha}{\bf v_{\alpha}})=0,
\end{equation}
and the induction equation
\begin{equation}
\label{ind} {\bf \nabla\times B} = \frac{1}{c}\frac{\partial {\bf
E}}{\partial t}+\frac{4\pi}{c}\sum_{\alpha=e^{\pm},b}{\bf
J_{\alpha}},
\end{equation}
where $n_{\alpha}$ and ${\bf J_{\alpha}}$ are the density and the
current, respectively.

Rewriting the Euler equation in Eq. (\ref{eul}) for the leading
state and taking into account the frozen-in condition ${\bf E}_0+
\frac{1}{c}{\bf v}_{0\alpha}\times{\bf B}_0=0$, the solution for
ultra relativistic particle velocities writes as follows (\cite{mr})
\begin{equation}
\label{vr} v^{0}_{\theta}\equiv v_{_\parallel} = c\,\cos(\Omega t +
\varphi),
\end{equation}
where $v_{_\parallel}$ denotes the velocity component along the
magnetic field lines and $\varphi$, the initial phase of each
particle.

The centrifugal force eventually causes the separation of charges in
plasma consisting of several species. This process becomes so
important that the corresponding electromagnetic field affects the
dynamics of the charged particles. Therefore, the produced electric
field should also be considered in Eq. (\ref{eul}) as the next term
to be approximated (\cite{mnras}).

To clarify the mathematical treatment, we linearize the set of Eqs.
(\ref{eul}-\ref{ind}), assuming that, to the zeroth approximation,
the flow has a longitudinal velocity satisfying Eq. (\ref{vr}) and
also drifts along the $x$-axis because of the curvature of magnetic
field lines (see Fig. \ref{fig3}):

\begin{equation}
\label{drift} u_{\alpha}= \frac{\gamma_{\alpha_0}
v_{_\parallel}^2}{\omega_{B_{\alpha}} R_B},
\end{equation}
where $u_{\alpha}$ is the drift velocity; $\omega_{B_{\alpha}} =
e_{\alpha}B_0/mc$, $R_B$ is the curvature radius of magnetic field
lines, and $B_0$ is the magnetic induction in the leading state. In
our case, the curvature drift contributes to ${\bf J}$ [see Eq.
(\ref{ind})] as a source of additional current.

We represent all physical quantities as the sum of the zeroth and
the first order terms
\begin{equation}
\label{expansion} \Psi\approx \Psi^0 + \Psi^1,
\end{equation}
where
$$\Psi\equiv \{n,{\bf v},{\bf p},{\bf E},{\bf B}\}.$$
We then express the perturbed quantities as
\begin{equation}
\label{pert} \Psi^1(t,{\bf r})\propto\Psi^1(t) \exp\left[i\left({\bf
kr} \right)\right] \,,
\end{equation}
examining only the $x$ components of Eqs. (\ref{eul},\ref{ind}),
considering the perturbations with $k_{\theta}\ll k_x$ and $k_r =
0$, and bearing in mind that $v^1_{r}\approx cE^1_x/B_{0}$, one can
easily reduce the set of Eqs. (\ref{eul}-\ref{ind}) to the form

\begin{equation}
\label{eulp} \frac{\partial p^1_{{\alpha}_x}}{\partial
t}-i(k_xu_{\alpha}+k_{\theta}v_{_\parallel})p^1_{{\alpha}_x}=
\frac{e_{\alpha}}{mc}v_{_\parallel}B^1_{r},
\end{equation}
\begin{equation}
\label{contp} \frac{\partial n^1_{\alpha}}{\partial
t}-i(k_xu_{\alpha}+k_{\theta}v_{_\parallel})n^1_{\alpha}=
ik_xn_{\alpha}^0v^1_{\alpha_x},
\end{equation}
\begin{equation}
\label{indp} -ik_{\theta}cB^1_{r} = 4\pi
\sum_{\alpha=e^{\pm},b}e_{\alpha}(n_{\alpha}^0v^1_{\alpha_x}+n_{\alpha}^1u_{\alpha}).
\end{equation}
Introducing a special ansatz for $v^1_{\alpha_x}$ and $n^1_{\alpha}$

\begin{equation}
\label{anzp} v^1_{\alpha_x}\equiv V_{\alpha_x}e^{i{\bf
kA_{\alpha}(t)}},
\end{equation}
\begin{equation}
\label{anzn} n^1_{\alpha}\equiv N_{\alpha}e^{i{\bf kA_{\alpha}}(t)},
\end{equation}
where
\begin{equation}
\label{Ax} A_{\alpha_x}(t) = \frac{u_{\alpha}}{2\Omega}\left(\Omega
t + \varphi\right) + \frac{u_{\alpha}}{4\Omega}\sin[2\left(\Omega t
+ \varphi\right)],
\end{equation}
\begin{equation}
\label{Af} A_{\alpha_\theta}(t) = \frac{c}{\Omega}\sin(\Omega t+
\varphi),
\end{equation}
and substituting Eqs. (\ref{anzp},\ref{anzn}) into Eqs.
(\ref{eulp},\ref{contp}), one derives the expressions

\begin{equation}
\label{vx} v^1_{{\alpha}_x} =
\frac{e_{\alpha}}{mc\gamma_{\alpha_0}}{\rm e}^{i{\bf
kA_{\alpha}}(t)}\int^t{\rm e}^{-i{\bf
kA_{\alpha}}(t')}v_{_\parallel}(t')B_{r}(t')dt',
\end{equation}

$$n^1_{\alpha} =
\frac{ie_{\alpha}n_{{\alpha}}^0k_x}{mc\gamma_{\alpha_0}}{\rm
e}^{i{\bf kA_{\alpha}}(t)}\int^tdt'\int^{t''}{\rm e}^{-i{\bf
kA_{\alpha}}(t'')}v_{_\parallel}(t'')B_{r}(t'')dt''.$$
\begin{equation}
\label{n}
\end{equation}
Combining Eqs. (\ref{vx},\ref{n}) with Eq. (\ref{indp}), it is
straightforward to reduce it to the form
$$ -ik_{\theta}cB^1_{r}(t)
=\sum_{\alpha=e^{\pm},b}\frac{\omega^2_{\alpha}}{\gamma_{\alpha_0}c}{\rm
e}^{i{\bf kA_{\alpha}}(t)}\int^t{\rm e}^{-i{\bf
kA_{\alpha}}(t')}v_{_\parallel}(t')B_{r}(t')dt'+ $$
\begin{equation}
\label{ind1}i\sum_{\alpha=e^{\pm},b}\frac{\omega^2_{\alpha}}{\gamma_{\alpha_0}c}k_xu_{\alpha}{\rm
e}^{i{\bf kA_{\alpha}}(t)}\int^tdt'\int^{t''}{\rm e}^{-i{\bf
kA_{\alpha}}(t'')}v_{_\parallel}(t'')B_{r}(t'')dt'',
\end{equation}
Where $\omega_{\alpha} = e\sqrt{4\pi n_{\alpha}^0/m}$ is the plasma
frequency. To simplify Eq. (\ref{ind1}), one may use the following
identity
\begin{equation}
\label{bess} {\rm e}^{\pm ix\sin y}=\sum_s J_s(x){\rm e}^{\pm isy},
\end{equation}
where $J_s(x)$ ($s=0;\pm 1;\pm 2 \ldots$) is the Bessel function of
integer order (\cite{abrsteg}). Eq. (\ref{ind1}) then reduces to the
form
$$B_{r}(\omega) =
-\sum_{\alpha=e^{\pm},b}\frac{\omega^2_{\alpha}}{2\gamma_{\alpha_0}k_{\theta}c}\sum_{\sigma
= \pm
1}\sum_{s,n,l,p}\frac{J_s(g_{\alpha})J_n(h)J_l(g_{\alpha})J_p(h)}{\omega
+ \frac{k_xu_{\alpha}}{2}+\Omega (2s+n) } \times$$ $$\times
B_{r}\left(\omega+\Omega
\left(2[s-l]+n-p+\sigma\right)\right)\left[1-\frac{k_xu_{\alpha}}{\omega
+ \frac{k_xu_{\alpha}}{2}+\Omega (2s+n)}\right]\times$$
$$\times {\rm e}^{i\varphi\left(2[s-l]+n-p+\sigma\right)}+$$
$$+\sum_{\alpha=e^{\pm},b}\frac{\omega^2_{\alpha}k_xu_{\alpha}}{4\gamma_{\alpha_0}k_{\theta}c}\sum_{\sigma,\mu
= \pm
1}\sum_{s,n,l,p}\frac{J_s(g_{\alpha})J_n(h)J_l(g_{\alpha})J_p(h)}{\left(\omega
+ \frac{k_xu_{\alpha}}{2}+\Omega (2[s+\mu]+n)\right)^2 } \times$$
\begin{equation}
\label{disp1} \times B_{r}\left(\omega+\Omega
\left(2[s-l+\mu]+n-p+\sigma\right)\right)\times {\rm
e}^{i\varphi\left(2[s-l+\mu]+n-p+\sigma\right)},
\end{equation}
where

$$g_{\alpha} = \frac{k_xu_{\alpha}}{4\Omega}, \;\;\;\;\;\;\;\;\;\;\;\;h =
\frac{k_{\theta}c}{\Omega}.$$

To solve Eq. (\ref{disp1}), we must examine similar equations, e.g.,
rewrite Eq. (\ref{disp1}) for $B_{r}(\omega\pm\Omega)$,
$B_{r}(\omega\pm 2\Omega)$, etc.. This means that one has to solve a
system with an infinite number of equations, which makes the problem
impossible to handle. Therefore, the only solution is to consider
the physics close to the resonance condition, which provides the
cutoff to the infinite row in Eq. (\ref{disp1}) making the problem
solvable (\cite{silin}).

Let us consider the resonance condition, which corresponds to the
curvature drift modes. As is clear from Eq. (\ref{disp1}), the
proper frequency for the CDI equals
\begin{equation}
\label{freq} \omega_0\approx -\frac{k_xu_{\alpha}}{2}.
\end{equation}
Therefore, physically meaningful solutions relate the case, when
$k_xu_{\alpha}/2<0$. The present condition implies that $2s+n = 0$
and $2[s+\mu]+n = 0$. On the other hand, it is easy to check that,
for the typical quantities of {\it $1$-second} pulsars,
$\gamma_b\sim 10^6$ and $\lambda\sim 6\times 10^{10}- 3\times
10^{11}cm$ (where $\lambda\approx\lambda_x=1/k_x$ is the
wavelength), one has $|k_x u_{\alpha}/2|\sim (0.02\div 0.1)s^{-1}$
(here, it is assumed that $k_x<0$ and $u_{b}>0$, otherwise the
resonance frequency would be negative).

Since particles have different phases, to solve Eq. (\ref{disp1}),
we must examine the average value of $B_r$ with respect to
$\varphi$. Then, by taking into account the formula

$$\frac{1}{2\pi}\int{\rm e}^{iN\varphi} d\varphi= \delta_{N,0},
$$
preserving only the leading terms of Eq. (\ref{disp1}), and also
taking into account that the beam components exceed the
corresponding plasma terms by many orders of magnitude, one derives
the dispersion relation for the CDI (\cite{mnras})

\begin{equation}
\label{disp} \left(\omega + \frac{k_xu_{b}}{2}\right)^2 \approx
\frac{3\omega^2_{b}k_xu_{b}}{2\gamma_{b_0}k_{\theta}c}\left[J_0\left(\frac{k_xu_{b}}
{4\Omega}\right)J_{0}\left(\frac{k_{\theta}c}{\Omega}\right)\right]^2.
\end{equation}

To determine the CDI growth rate, $\Gamma$, let us write
$\omega\equiv\omega_0+i{\Gamma}$ and substitute this into Eq.
(\ref{disp}). Then, it is easy to show that the increment is given
by

\begin{equation}
\label{increm} \Gamma \approx
\left(-\frac{3}{2}\frac{\omega^2_{b}}{\gamma_{b_0}}\frac{k_xu_{b}}{k_{\theta}c}\right)
^{1/2}\left|J_0\left(\frac{k_xu_{b}}
{4\Omega}\right)J_{0}\left(\frac{k_{\theta}c}{\Omega}\right)\right|.
\end{equation}
We can qualitatively analyze how the shape of magnetic field lines
changes with time. After perturbing the magnetic field in the
transverse direction, the toroidal component will strengthen and the
field lines will gradually lag behind the rotation. This process
will inevitably influence the dynamics of the particles. As a
result, the acceleration will be lower than in the case of straight
field lines. Since the CDI is centrifugally excited, the
corresponding efficiency will also decrease, completely vanishing
for such a configuration of the magnetic field lines, when the
particles do not accelerate centrifugally at all.

Let us assume that the corresponding critical value of the toroidal
component, when the instability diminishes, is $B_{r}$. Then, from
simple considerations, we can estimate the corresponding timescale
of this transformation. Indeed, referring to Fig. \ref{fig3}b,
$\tan\chi = B_{r}/B_{\theta}$. On the other hand, the toroidal
component of magnetic field behaves with time as
$$\label{br} B_{r}\approx B^0_r e^{\Gamma t},$$ where $B^0_r$ denotes
the initial perturbed value of the toroidal component. Consequently,
the timescale of the process can be estimated to be
\begin{equation}
\label{time} T \approx - \frac{1}{\Gamma(\lambda_x)}\ln
\left(\frac{B^0_r}{B_{\theta}\tan\chi} \right).
\end{equation}

\begin{figure}
 \par\noindent
 {\begin{minipage}[t]{1.\linewidth}
 \includegraphics[width=\textwidth] {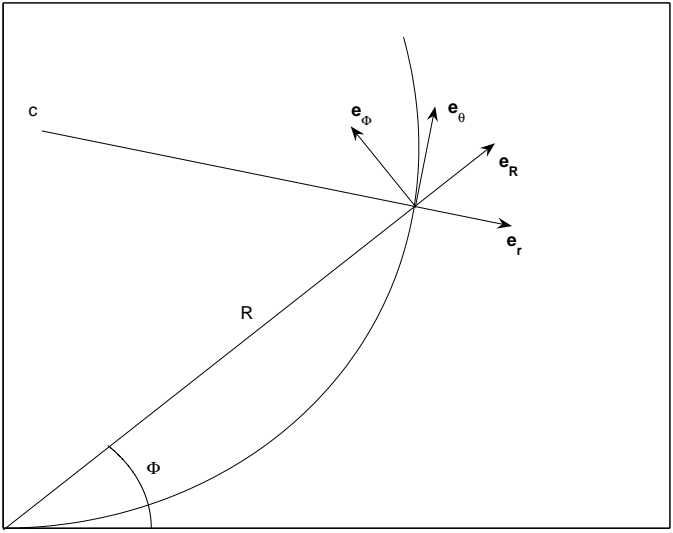}
 \end{minipage}
 }
 \caption[ ] {The arm of Archimedes
spiral in polar coordinates ($\Phi,R$). Two orthonormal bases are
considered: i) polar components of unit vectors, (${\bf e}_\Phi,
{\bf e}_R$); and ii) normal and tangential components of unit
vectors, (${\bf e}_r, {\bf e}_{\theta}$), respectively. $C$ is the
center of the curvature. }\label{fig}
 \end{figure}

%
\section{Kinematics of particles moving along the curved magnetic field lines}
%

We describe the kinematics of particles moving along the corotating
curved magnetic field lines. In particular, we adopt the Archimedes'
spiral for the configuration of magnetic field lines
\begin{equation}
\label{spir} \Phi = aR,
\end{equation}
where $\Phi$ and $R$ are the polar coordinates and $a = {\it const}$
(see Fig. \ref{fig}). Rogava et al. (2003) showed that the dynamics
of the particle motion asymptotically tends to the force-free
regime, if the particle slides along the rotating channel that has
the shape of the Archimedes' spiral. However, in the laboratory
frame (LF), the particle follows straight linear paths with constant
velocities. In this case, an observer from the LF will measure the
effective angular velocity
\begin{equation}
\label{omef} \Omega_{ef} = \Omega + \frac{d\Phi}{dt} =\Omega + av,
\end{equation}
where $v$ is the radial velocity of a particle motion, and $\Omega$
is the angular velocity of rotation. If the particle moves without
acceleration in the LF, the corresponding effective angular velocity
must be equal to zero ($\Omega_{eff}=0$). In this case, Eq.
(\ref{omef}) infers that
\begin{equation}
\label{spir1} v = v_c \equiv -\frac{\Omega}{a},
\end{equation}
which means that, for any Archimedes' spiral with $a<-\Omega/c$, the
LF trajectory of the particle becomes a straight line if it moves
with a certain "characteristic velocity", $v_c$.
\\*
The relativistic momentum of the particle is given by
\begin{equation}
\label{pr} P_{R} = \gamma mv,
\end{equation}
\begin{equation}
\label{pfi} P_{\Phi} = \gamma mR\Omega_{ef}\,,
\end{equation}
where $m$ and $\gamma$ are the rest mass and the Lorentz factor of
the particle, respectively. The expansion of the equation of motion,
$d{\bf P}/dt = {\bf F}$ (${\bf F}$ is the reaction force), in terms
of the radial component
\begin{equation}
\label{fr}F_{R} = -{{aR}\over{\sqrt{1+a^2R^2}}}|{\bf F}|,
\end{equation}
and the angular component
\begin{equation}
\label{ffi}F_{\Phi} = {{1}\over{\sqrt{1+a^2R^2}}}|{\bf F}|,
\end{equation}
of the reaction force, yields the equation (\cite{r03})

\begin{equation}
\label{a_r}\frac{d^2R}{dt^2}=\frac{\Omega-\gamma^2v(a + \Omega v
/c^2 )}{\gamma^2\kappa^2}\Omega_{ef}R,
\end{equation}
where
\begin{figure}
 \par\noindent
 {\begin{minipage}[t]{1.\linewidth}
 \includegraphics[width=\textwidth] {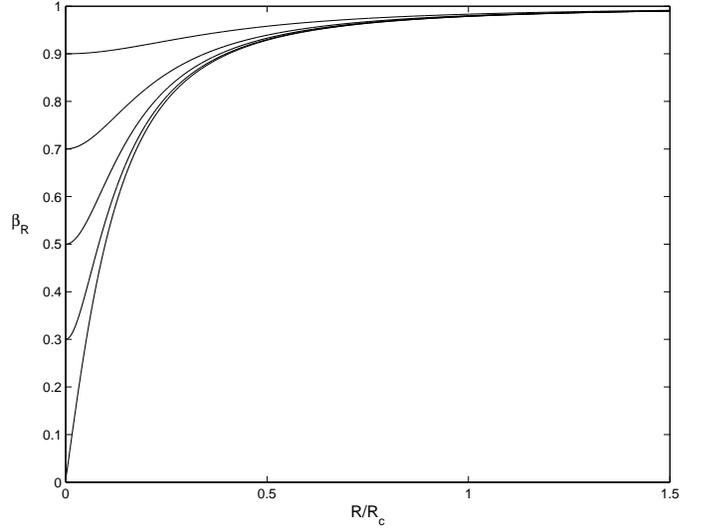}
 \end{minipage}
 }
 \caption[ ] {Behaviour of $\beta_R$
versus $R/R_c$. The set of parameters is $P = 1s$, $\beta_{R0} =
\{0.01; 0.3; 0.5; 0.7; 0.9\}$.}\label{fig1}
 \end{figure}
$$\kappa\equiv \left(1- \frac{\Omega^2 R^2}{c^2}
+a^2R^2\right)^{1/2}.$$ We note that $(d^2R/dt^2)\equiv 0$, when
$v=v_c=-\Omega/a$.

Since we are interested in relativistic flows, let us consider $v_c
= c$ setting $a = -\Omega /c$. In Fig. \ref{fig1}, we plot the
solution to Eq. (\ref{a_r}), that is the dependence of
$\beta_R\equiv v/c$ versus $R/R_c$ ($R_c$ is the light cylinder
radius) for the different initial values of $\beta_{R0}$. As is
clear from these plots, even though the initial velocity of
particles is weakly relativistic (e.g., $\beta_{R0} = 0.01$), it
asymptotically converges to the characteristic velocity, $v_c$. As a
result, the particle trajectory in the LF must become linear.

\begin{figure}
 \par\noindent
 {\begin{minipage}[]{0.35\linewidth}
 \includegraphics[width=\textwidth] {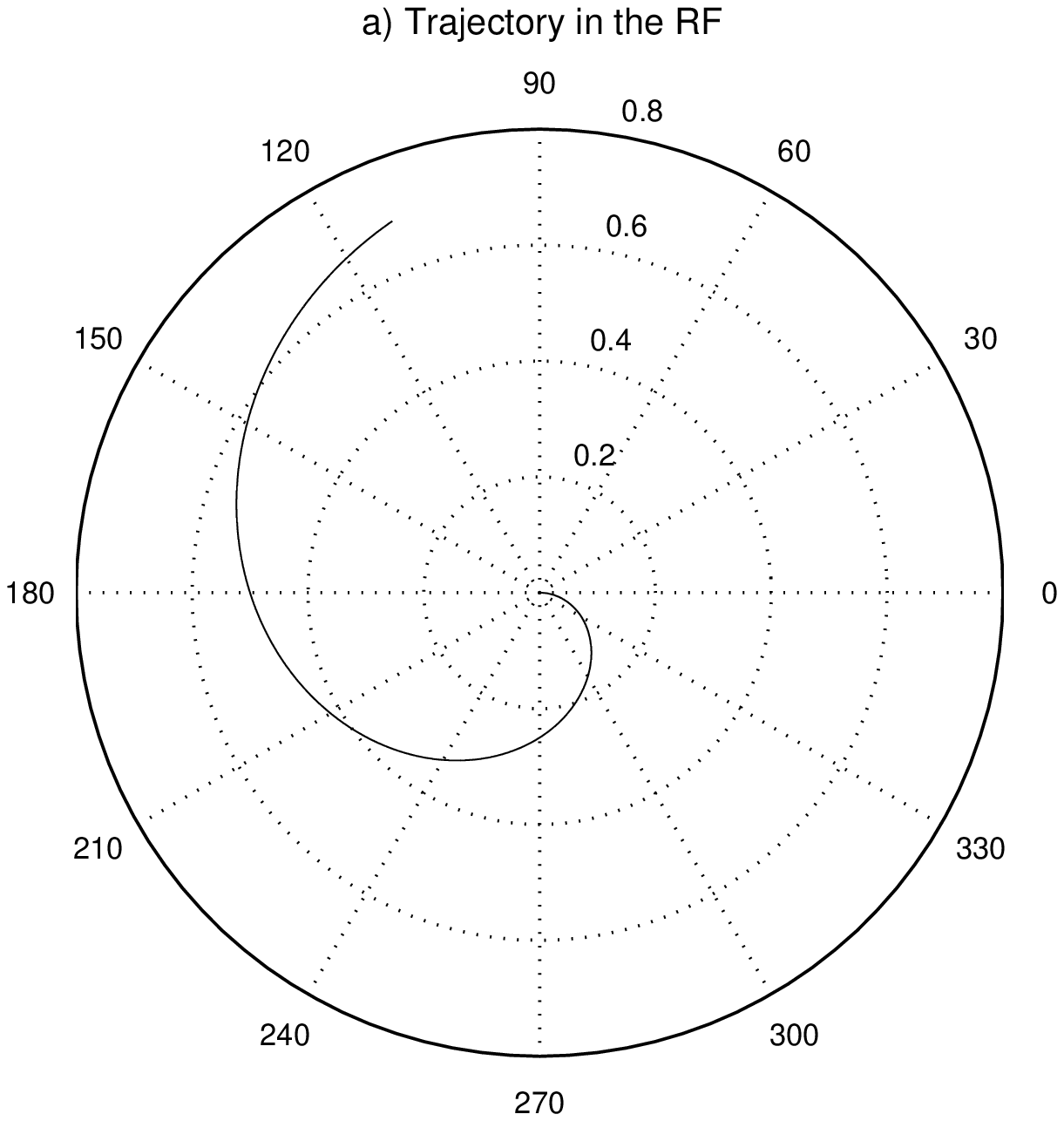}
 \end{minipage}
 }
 \hfill
 {\begin{minipage}[]{0.35\linewidth}
 \includegraphics[width=\textwidth] {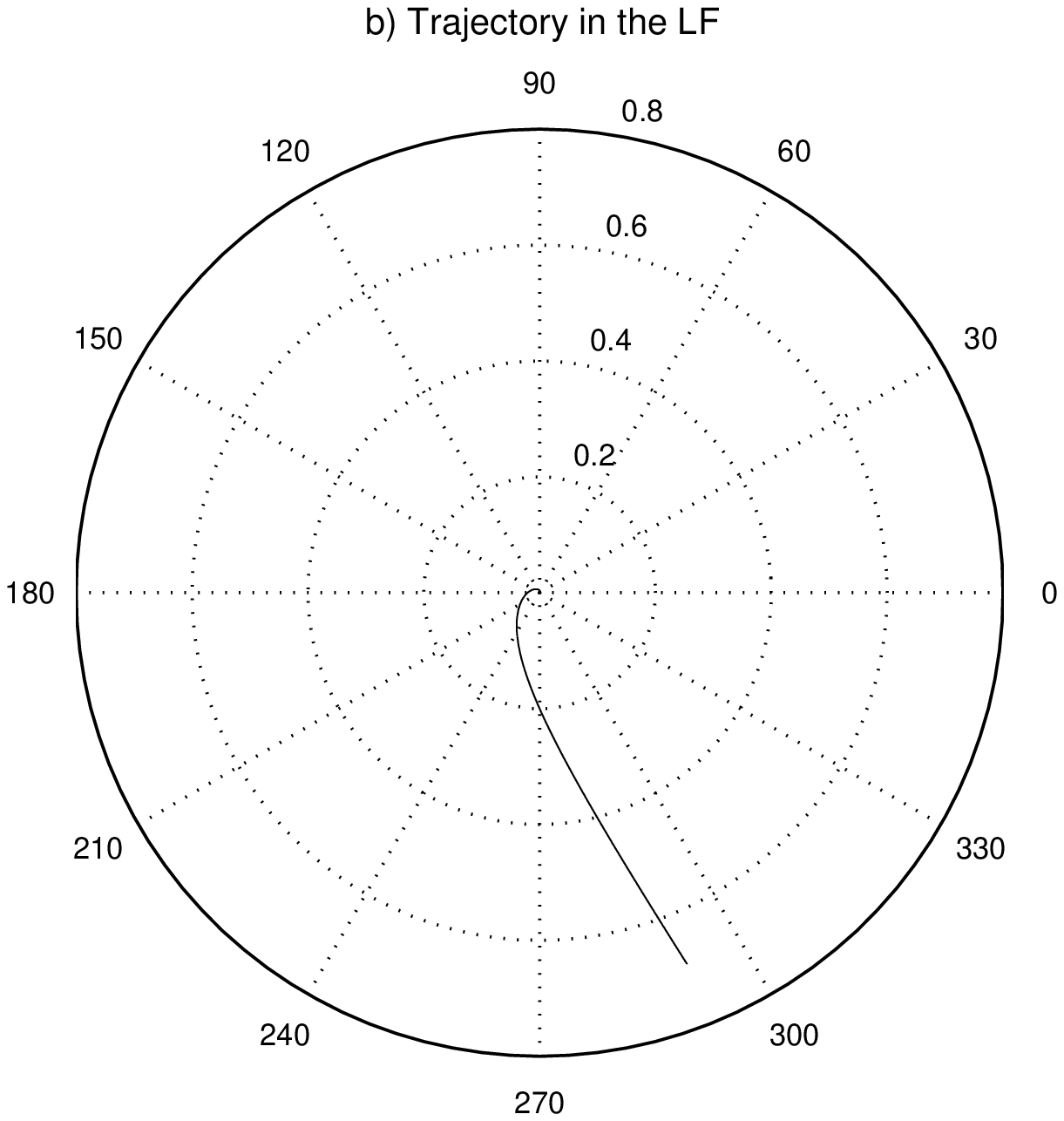}
 \end{minipage}
 }
 \caption[ ]{\footnotesize $a$) The trajectory of a
particle in the Rotational Frame (RF) of reference; and $b$) the
trajectory of a particle in the Laboratory Frame (LF) of reference.
The set of parameters is $P = 1s$ and $\beta_{R0}^{1} = 0.01$. The
radial distances are taken in terms of the light cylinder radius,
$R/R_c$. The trajectory of the particle in RF follows the
Archimedes' spiral ($a$), while the trajectory in the LF
asymptotically tends to a straight line configuration
($b$).}\label{fig2}
 \end{figure}

Indeed, in Fig. \ref{fig2}, we show the particle trajectory in the
Rotational Frame (RF) of reference (see Fig. \ref{fig2}a) as well as
in the LF (see Fig. \ref{fig2}b) for both $\beta_{R0} = 0.01$ and
the same spiral configuration with $a = -\Omega /c$. Observing the
particle trajectory from the LF, one can note that the path
asymptotically becomes linear, indicating that the rotational energy
pumping process diminishes. Therefore, we conclude that the magnetic
field with the of the Archimedes' spiral may guarantee the
saturation process of the CDI.

On the other hand, the equation of Archimedes' spiral yields
$\tan\chi=aR=B_r/B_{\theta}$ (see Fig. \ref{fig3}b), resulting in
$B_r=B_{\theta}$ at the LC when $a = -\Omega /c$.

\section{Discussion} \label{sec:discus}

\begin{figure}
 \par\noindent
 {\begin{minipage}{0.48\linewidth}
 \includegraphics[width=\textwidth] {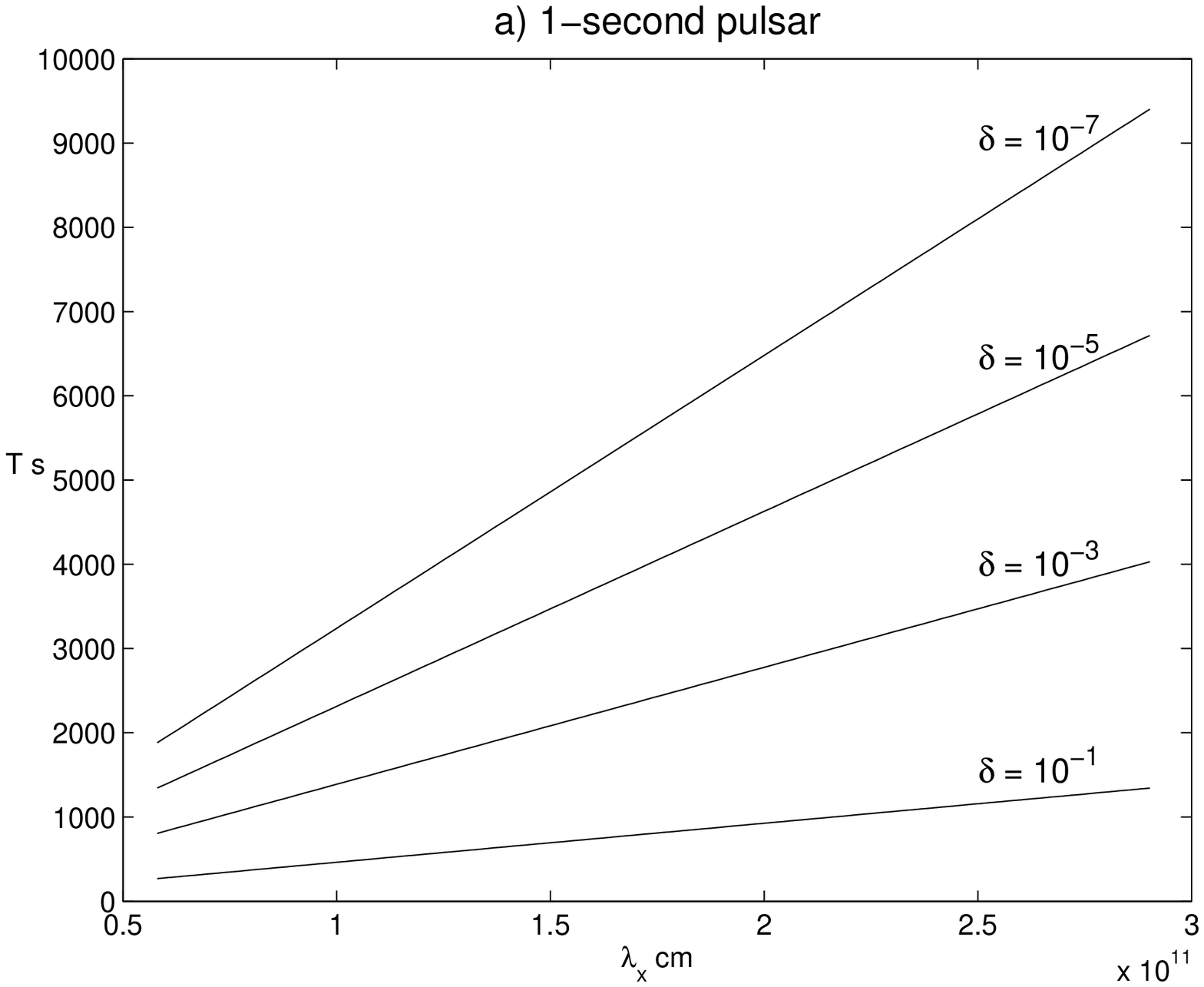}
 \end{minipage}
 }
 \hfill
 {\begin{minipage}{0.48\linewidth}
 \includegraphics[width=\textwidth] {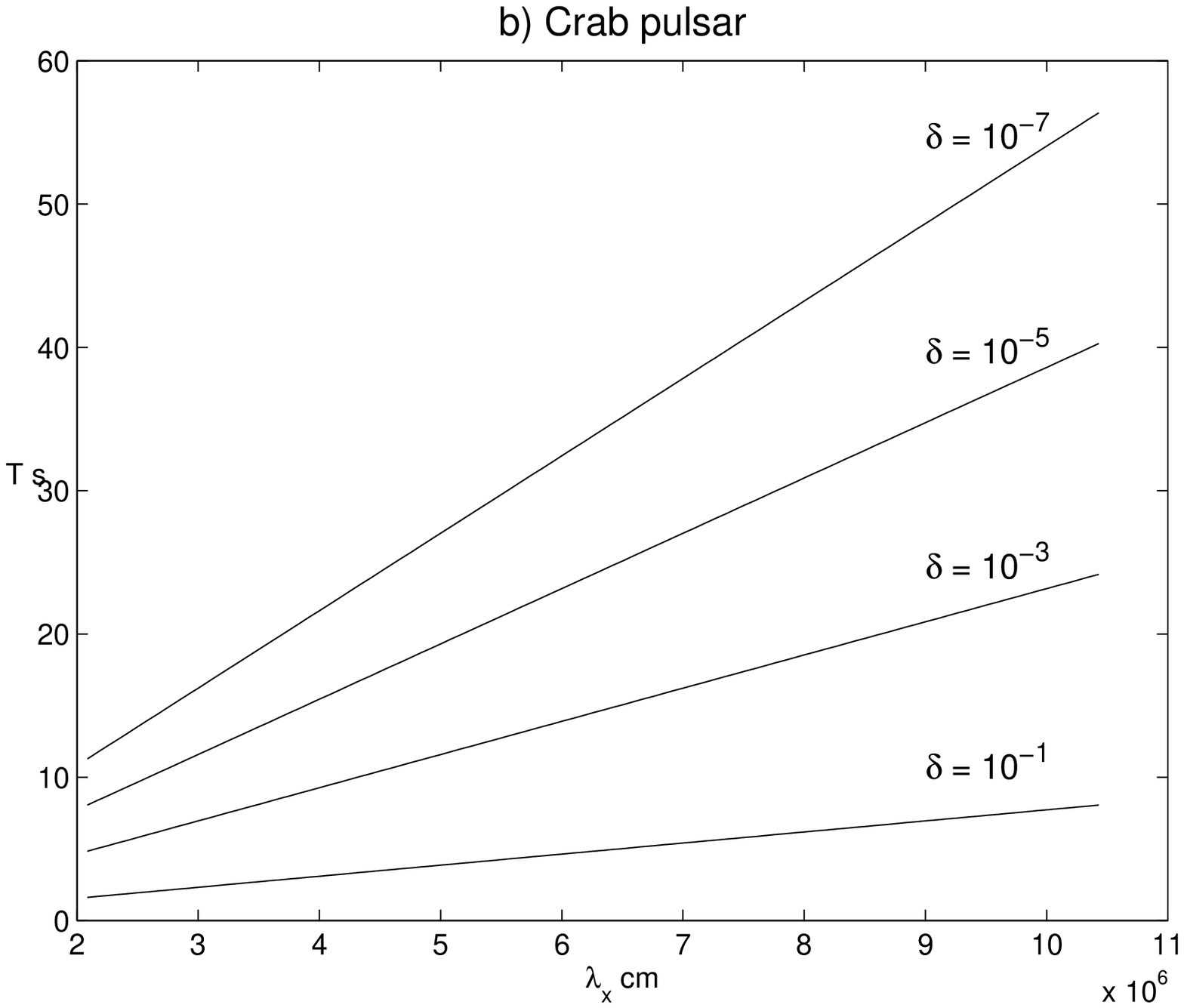}
 \end{minipage}
 }
 \caption[ ] {\footnotesize Here we show the
behaviour of the timescale of sweepback of the magnetic field lines
versus the wave length, $\lambda\approx\lambda_x$, for the {\it
1-second} pulsar [see (a)] and for the Crab pulsar, $P\approx
0.033s$ [see (b)], respectively. The set of parameters is: $\delta =
\{10^{-1}; 10^{-3}; 10^{-5}; 10^{-7}\}$, $\gamma_b\sim 10^6$ and
$\lambda_{\theta} = 1000R_{c}$.}\label{fig4}
 \end{figure}

We consider Eq. (\ref{time}) and plot the timescale of the
instability versus the wavelength, $\lambda_x$, for several values
of initial toroidal perturbations $B^0_r$. In Fig. \ref{fig4}, we
display the behavior of $T(\lambda_x)$, when $\tan\chi=1$, for
several values of initial perturbation
$B^0_r/B_{\theta}\equiv\delta\in \{10^{-1}; 10^{-3}; 10^{-5};
10^{-7}\}$. Two major applications are examined: (a) {\it 1-second}
pulsar and (b) the Crab pulsar. From Eq. (\ref{time}), we can infer
that the timescale is a continuously decreasing function of the
initial perturbation, $B^0_r$: as becomes smaller the $\delta$
parameter, so does the initial perturbation and, consequently, the
magnetic field lines need more time to achieve the required
structure. As observations show, the ratio $P/\dot{P}$ ranges from
$10^{11}s$ (PSR 0531 - Crab pulsar) to $10^{18}s$ (PSR 1952+29).
However, the greatest values of the twisting timescales, shown in
Fig. \ref{fig4}, vary between $10^4s$ and $10^2s$, which are shorter
by many orders of magnitude than $P/\dot{P}$, illustrating the high
efficiency of the instability.

The reconstruction of the magnetic field requires a certain amount
of energy, therefore it is essential to estimate pulsar's slowdown
luminosity ($L_p$) and compare it to the, so-called, "magnetic
luminosity" ($L_m\equiv \triangle E_m/T$, where $\triangle E_m$ is
the variation in the magnetic field energy due to the instability).
For $L_p$, one has

\begin{equation}
\label{rot1} L_{p} = I\Omega\dot{\Omega} =
I\frac{4\pi^2}{P^2}\frac{\dot{P}}{P},
\end{equation}
where $I\sim MR_p^2$ is the moment of inertia of the pulsar, and
$M\sim M_{\odot}\approx 2\times 10^{33}g$ and $R_p\sim 10^6cm$ are
the pulsar's mass and radius, respectively. For typical pulsars,
$\dot{P}/P\sim 10^{-15}s^{-1}$, the slowdown luminosity is estimated
to be
\begin{equation}
\label{rot2} L_{p}\approx 7.9\times 10^{31}erg/s.
\end{equation}
The "magnetic luminosity", $L_m$, can be estimated straightforwardly
\begin{equation}
\label{mag1} L_{m} = \frac{B_{r}^2}{4\pi T}\triangle V,
\end{equation}
where $\triangle V\sim R_c^2\triangle r=R_c^3 \varepsilon$
($\varepsilon\equiv\triangle r/R_c<<1$) is the volume, where the
process of twisting takes place. Since we are studying the
instability in the nearby zone of the LC, the magnetic field
components must be given as $B^0_r\sim \delta\cdot
B_{\theta}=\delta\cdot B_{p}(R_p/R_c)^3$, where $B_{p}\sim 10^{12}G$
is the magnetic field near the pulsar's surface. Substituting all
quantities into Eq. (\ref{mag1}) corresponding to the energy gain
for $\lambda_x\sim 2.5\times 10^{11}cm$, and $T\approx 10^3s$, (see
Fig. \ref{fig4}a), and bearing in mind Eqs. (\ref{br},\ref{time}),
one can show that for $\varepsilon\sim 0.1$ and $\delta\sim 0.1$ the
value of the "magnetic luminosity" is approximately equal to

\begin{equation}
\label{mag2} L_{m} \approx 7.3\times 10^{25}erg/s.
\end{equation}
A direct comparison between Eqs. (\ref{rot2}) and (\ref{mag2})
infers that $L_{m}<<L_{p}$, meaning that approximately only
$0.0001\%$ of the total energy budget goes to the reconstruction of
the magnetic field lines.

In Fig. \ref{fig4}b, we show the behavior in $T(\lambda_x)$ for the
Crab pulsar in the different cases of initial magnetic
perturbations. The dependence does not change qualitatively but the
twisting process changes quantitatively as the corresponding
timescale is now of the order of $\sim 10\d-10^2\,s$. The "magnetic
luminosity" of the Crab pulsar for $\delta = 0.1$, $\lambda_x =
7\times 10^6cm$, $T\approx 4\,s$ (see Fig. \ref{fig4}b), and
$\varepsilon = 0.1$ approximately equals [see Eq. (\ref{mag1})]
\begin{equation}
\label{cmag2} L_{m}^{Crab} \approx 4.7\times 10^{32}erg/s.
\end{equation}
On the other hand, a direct calculation of the Crab pulsar
($P\approx 0.033\,s$) luminosity, estimated by Eq. (\ref{rot1}),
yields
\begin{equation}
\label{crot2} L_{p}^{Crab}\approx 9.3\times 10^{38}erg/s.
\end{equation}
Therefore, as in the case of the Crab pulsar, the energy required to
reconstruct the magnetosphere averages $\sim 0.00005\%$ of the
pulsar's energy and the twisting process becomes feasible in this
case as well.

The sweepback mechanism described in this paper, therefore, appears
to be extremely efficient for pulsar magnetospheres. The curvature
drift instability may lead to the reconstruction of the magnetic
field lines in such a way that the dynamics of magnetosphere becomes
force-free, which in its turn, completely disrupts the instability,
thus saturating the CDI.

\section{Summary}\label{sec:summary}

      We have analytically examined the non-stationary pattern of the
      pulsar magnetosphere near the LC zone. The
      GJ current can be estimated
      by $J_{GJ} = en_b c$, giving the value of the corresponding magnetic field,
      $B_r\approx 4\pi J_{GJ}R_n/c$, where $R_n\approx B(R)/B'(R) = R/3$ (where $B'\equiv dB/dR$)
      is the length scale of the spatial
      inhomogeneity of the magnetic field. If we assume a dipolar configuration,
      then, taking the value of the GJ density, $n_b\approx \Omega B/(2\pi ec)$,
      into account, one can show that the toroidal magnetic field equals $\frac{2R}{3R_c}B$.
      Inside the light cylinder ($R< R_c$), this value is less than the required
      one-$B$ and reaches its maximum value, $\frac{2}{3}B$, on the light surface.
      Therefore, the GJ current cannot significantly change
      the configuration of the magnetic
      field. This current evidently exceeds the
      curvature drift current, because $J_{GJ}/J_{cur}\approx c/u_b\gg 1$.
      However, the drift current, $J_{cur}$, is not the source of the toroidal component, $B_r$,
      of the magnetic field, but it is a trigger mechanism for generation of
      the perturbed current, $J_1= e(n_b^0v^1_{b_x}+n_b^1u_b)$  responsible for the creation of $B_r$ (see Eq. (\ref{indp})).
      The source of the instability of the current and the resulting
      magnetic field is the pulsar's rotational energy. We have found that the instability
      is achieved by the parametric mechanism, which effectively pumps the pulsar
      rotational energy directly into the generated mode.
      The process lasts until the plasma dynamics reaches the force-free regime of motion and the
      overall magnetosphere relaxes to the steady state configuration.
      \cite{buc} considered the dynamics of rotating pulsar winds by
      performing the numerical solution of relativistic
      magnetohydrodynamic (RMHD) equations in the Schwarzschild metric. The system was allowed to relax to a steady
      state configuration rapidly approaching the force-free regime.
      We suppose that the full set of RMHD equations comprises the terms applicable to CDI.
      Therefore, we highlight the numerical results obtained by \cite{buc} as
      confirmation of our theory. On the other hand, the aim of the present
      work was to explain the saturation of the CDI in
      terms of the generation of currents, which makes the physics of
      the process more transparent.

      The main aspects of the present work can be summarized as follows:

\begin{enumerate}
      \item Examining the pulsar magnetospheric relativistic plasma, we have studied
      the role of the parametrically excited CDI in the process
      of sweepback of magnetic field lines and the saturation process of the
      instability. The present parametric instability is based on the
      method developed by Silin (1973), but differs from this in
      principal, since we study an alternating centrifugal force
      instead of an alternating electric force.

      \item The linear analysis of the Euler, continuity, and induction
      equations yields the dispersion relation governing
      the CDI.

      \item Considering the resonance frequencies of the sweepback process,
      an expression of the instability increment has been
      obtained.

      \item On the basis of the expression of the instability growth rate,
      we have derived the formula of the transition timescale of quasi-linear
      configuration of field lines into the
      Archimedes' spiral. The particles' motion is force-free along these magnetic field lines,
      leading to the saturation of the instability.

      \item The transition timescale has been compared to
      the wavelength for {\it 1-second} pulsars and the Crab pulsar.
      For both cases it was shown that the corresponding timescale is
      shorter than pulsar's spin down rates by many orders of
      magnitude illustrating the high efficiency of the discussed process.

      \end{enumerate}

\section*{Acknowledgments}

The authors are grateful to Dr. N. Bucciantini for interesting
discussions. Z.O. and D.Sh. acknowledge the hospitality of the Abdus
Salam International Centre for Theoretical Physics (Trieste, Italy).
The research was supported by the Georgian National Science
Foundation grant GNSF/ST06/4-096.


\begin{thebibliography}{999}
\bibitem[Abramovitz \& Stegan 1965]{abrsteg}Abramovitz, M. \&
Stegan, I., 1965, Handbook of Mathematical Functions, (eds.: Dover
Publications Inc.: New York), p. 320
\bibitem[Bucciantini et al. (2006)]{buc} Bucciantini N., Thompson T.A.,
Arons J., Quataert E. \& Del Zanna L., 2006, MNRAS, 368, 1717
\bibitem[Galeev \& Sagdeev 1973]{gal} Galeev \& Sagdeev, 1973, Nucl. Fussion, 13,
603
\bibitem[Goldreich \& Julian 1969]{gj} Goldreich, P. \& Julian, W.H., 1969, ApJ, 157, 869
\bibitem[Kazbegi et al. 1989]{kmm89} Kazbegi A.Z., Machabeli G.Z. \& Melikidze G.I., in {\it Joint Varenna-Abastumani International School} \& {\it Workshop on Plasma Astrophysics}, volume ESA SP-285, edited by T.D.
Guyenne, (European Space Agency, Paris, 1989), p. 277
\bibitem[Kazbegi et al. 1991a]{kmm91a}  Kazbegi A.Z., Machabeli G.Z. \& Melikidze G.I., 1991a, MNRAS, 253, 377
\bibitem[Kazbegi et al. 1991b]{kmm91b}  Kazbegi A.Z., Machabeli G.Z. \& Melikidze G.I., 1991b,
Aust. J. Phys. 44, 573
\bibitem[Machabeli et al. 2000]{mmsh2000} Machabeli G.Z., Mchedlishvili G.Z., \& Shapakidze D.E., 2000, Ap\&SS, 271, 277
\bibitem[Machabeli et al. 2005]{incr1} Machabeli G., Osmanov Z. \& Mahajan S., 2005, Phys. Plasmas 12,
062901
\bibitem[Machabeli \& Rogava 1994]{mr} Machabeli, G.Z. \& Rogava, A. D., 1994, Phys.Rev. A, 50,
98
\bibitem[Manchester \& Taylor 1977]{manch} Manchester R. N. \& Taylor
J. H., 1977, Pulsars (ed.: W. H. Freeman and Company: San Francisco)
\bibitem[Max 1973]{max} Max C., 1973, Phys. Fluids, 16, 1480
\bibitem[(1994)]{mestshib} Mestel L. \& Shibata S., 1994, MNRAS, 271, 621
\bibitem[Osmanov et al. 2008]{mnras} Osmanov, Z., Dalakishvili, Z. \& Machabeli, Z. 2008, MNRAS, 383, 1007
\bibitem[Piddington (1957)]{pid} Piddington J.H., 1957, AuJPh, 10, 530
\bibitem[Rogava et al. 2003]{r03} Rogava, A. D., Dalakishvili, G. \& Osmanov, Z.N., 2003, Gen. Rel. and Grav. 35, 1133
\bibitem[Shapakidze et. al 2003]{shmmkh03} Shapakidze, D., Machabeli, G. Melikidze, G., \& Khechinashvili, D., 2003,
Phys.Rev.E, 67, 026407
\bibitem[Silin \& Tikhonchuk 1970]{silin} Silin V.P. \& Tikhonchuk V.T., 1970, J. Appl. Mech. Tech. Phys., 11, 922
\bibitem[Silin 1973]{silin1} Silin V.P., 1973, 'Parametricheskoe Vozdeistvie izluchenija
bol'shoj mosshnosti na plazmu', Nauka, Moskva
\bibitem[Weber \& Davis 1967]{weber} Weber E.J. \& Davis L.Jr., 1967, \apj, 148, 217


\end{thebibliography}
\end{document}